# A collection of database industrial techniques and optimization approaches of database operations

Catapang, Jasper Kyle

Databases play an essential role in our society today. Databases are embedded in sectors like corporations, institutions, and government organizations, among others. These databases are used for our video and audio streaming platforms, social gaming, finances, cloud storage, e-commerce, healthcare, economy, etc. It is therefore imperative that we learn on how to properly execute database operations and efficiently implement methodologies so that we may optimize the performance of databases.

## Terminologies

Database - a structured set of data that is accessible in many ways through a computer

Data mart - a more specified data warehouse whose data are obtained by selecting and summarizing data from the data warehouse [1]

Data warehouse - a subject-oriented, integrated, time-variant, nonvolatile collection of data used in support of management decision-making processes [2]

Feature/field - a column in the table of a database

Grain - the length of time associated with each record in the table [3]

Joining - process of combining data from various sources into a single table or view [4]

Record - a row or entry in the table of a database

Schema - a logical container of tables, views, and stored procedures

Table - a set of values using a model of vertical columns and horizontal rows

View - a result set of a stored query on the data

# Overview

A database is a usually large collection of data organized especially for rapid search and retrieval [5]. Creation, retrieval, modification, and deletion are the four basic database operations. These four functions are more commonly termed as create, read, update, and delete, or CRUD.

1. Creation is the addition of new entries or records onto a table in the database.
2. Retrieval is the searching or viewing of existing records in a table in the database.
3. Modification is the editing of existing records in a table in the database for the following purposes:
     a. content error - value is erroneous (typographical error, false or deceptive data)
     b. content update - value is replaced due to changes over a period of time
4. Deletion is the removal of existing records in a table in the database.

# Transient data versus periodic data

Before the approaches and techniques are discussed, it is necessary to introduce the concepts of transient data and periodic data. Transient data are data overwritten on top of previous records or data that are permanently erased when modification and deletion are executed respectively—effectively destroying the history of the data. Periodic data are data that are never physically altered or deleted. The use of these two types of data depend on the nature and philosophy of the industry. (Salisbury, D., 2007).

# Database operation approaches

## Creation

1. **Successive addition**

Successive addition is the addition of records onto the table one at a time. This is efficient when the addition of new data does not occur frequently and that the records to be added are few—a record count that is preferably countable by the human hands instantly (less than or equal to 10).

Successive addition is inefficient when a) record addition is getting more frequent in the number of times it is performed , or b) the record addition occurs at an interval (hourly, daily, every n days, etc.). (Vanga, S., 2011).

2. **Bulk addition**

Bulk addition is the addition of records onto the table as one whole batch. Bulk approach to record addition is highly effective when a) data is huge but manageable (hundreds to thousands of records per execution), b) the database is locally stored, and that c) the machine that the database operation is done is sufficient in terms of machine requirements for huge data processing.

This is not ideal for huge, overwhelming data (thousands to millions of records per record addition), for online databases, as it would take a significant amount of time to upload these records, and for machines that do not meet huge data processing requirements as it would also slow down performance.

3. **Partitioned addition**

Partitioned addition is the addition of records onto the table by grouping data into partitions or batches. With partitioned addition, the overwhelming nature of a thousand to million record data would be reduced and the heavy load would be addressed if the database is stored online. It would also relieve some computational intensity on substandard machines.

The number of partitions is the approach's strength and weakness. Using too many partitions would result to something like using the successive addition approach. On the other hand, using a few partitions is like performing bulk addition repetitively.

4. **In-parallel addition**

In-parallel addition is the addition of records onto the table by executing bulk addition or partitioned addition in-parallel using multithreading. (Shyamsundar, A., 2016).

# Retrieval
## A. Retrieval of records in a date range
1. **First occurrence approach**

First occurrence approach is the retrieval of records from the table by searching the first occurrence of the start date from the table and keep on selecting the records to include until the first occurrence of the date that is beyond the end date.

2. **Granular lookup schema approach (GLS approach)**

Granular lookup schema approach is the retrieval of records from table A by using a lookup table B provided that the database is structured such that an extra column(s) for labeled grains exist(s). The label would be counterchecked at lookup table B. This would significantly reduce search time because records are labeled in grains instead of individual dates per records.

Example:

1) Customer table with first occurrence approach

Scenario: Select date from August 1, 2018 to August 2, 2018

| Transaction number | Customer ID | Customer Name | Subscription | Date |
|---|---|---|---|---|
| 1 | RSoLGFtpbJ | Anne Allgood | AMZ20 | 07-05-18 |
| 2 | 7FbkoTnEMh | Barbie Banderas | AMZ30 | 07-05-18 |
| ... | ... | ... | ... | ... |
| 534 | s0E78xxKuG | Catherine Calvin | AMZ20 | 07-31-18 |
| 535 | IrmfSP9ZjJ | Diana Delevingne | AMZ20 | 08-01-18 |
| 536 | CqHVuthBnH | Elise Everett | AMZ40 | 08-02-18 |

The first occurrence approach had to go through the records from transaction number (TN) 1 to TN 534 before reaching TN 535 which is the start date.

2) Adding a lookup table (GLS approach)

Scenario: Select date from August 1, 2018 to August 2, 2018

| Code (column with arbitrary codes) | Starting transaction number (indices) |
|---|---|
| Q3M7 (Q3M7 = $3^{rd}$ quarter, month 7) | 1 |
| Q3M8 (Q3M8 = $3^{rd}$ quarter, month 8) | 535 |
| ... | ... |

The GLS approach only had to look up for Q3M8 for August 1, 2018 and only performed a go-to instead of a linear search. The granularity of the lookup table depends on database engineer implementation preference. By introducing codes to represent periods of time, we shorten the searching time.

## B. Retrieval of records for a specific entity

### 1. Exhaustive search approach for retrieval (ESR approach)

Exhaustive search approach is the retrieval of records by searching the first occurrence of a specific entity (person, product, service, location) and keep on selecting records that match until the entire table is searched. (LibGuides, n.d.).

### 2. Indexed entity approach for retrieval (IER approach)

Similar to the GLS approach, a lookup table is created containing the identifiers of all the entities in the main table. Each record in the main table is indexed and similar names have the indices aggregated in the lookup table.

Example:

Indexed patient table

| Patient id | Case number indices |
|---|---|
| 28435300131710927001 | 3, 5, 6, 11, 18 |
| 56729845935643075507 | 2, 4, 10 |
| ... | ... |

The indexed entity approach already has the case number indices that correspond to the patient. If individual case number indices are desired per row, then a simple *row explosion* would be done.

A row explosion is the transformation of a linear record into its individual indices.

Example:

| 28435300131710927001 | 3, 5, 6, 11, 18 |
|---|---|

The record above when performed row explosion on would be:

| 28435300131710927001 | 3 |
|---|---|
| 28435300131710927001 | 5 |
| 28435300131710927001 | 6 |
| 28435300131710927001 | 11 |
| 28435300131710927001 | 18 |

# Modification

## A. Maintenance of historical data

### 1. Transient modification

Transient modification is the editing of data over the previous content. This is used when history is not important according to the database engineer.

### 2. Periodic modification

Periodic modification is the addition of data instead of the actual modification of data and is labeled the updated value. The only value modified is the marker column indicating it is the current value.

Example:

Online assignment submission table

| Student name | Submitted | Timestamp | Currency |
|---|---|---|---|
| Cruz, Michael T. | None | NULL | Current |

And when the assignment was submitted and resubmitted over and over,

| Student name    | Submitted   | Timestamp         | Currency |
|-----------------|-------------|-------------------|----------|
| Cruz, Michael T.| None        | NULL              | NULL     |
| Cruz, Michael T.| file(1).txt | 08-23-18 23:58:40 | NULL     |
| Cruz, Michael T.| file(2).txt | 08-23-18 23:59:54 | Current  |

Here, it shows the multiple attempts of Michael to submit an assignment.

## B. Updating based on key term

### 1. Exhaustive search approach for modification (ESM approach)

Exhaustive search approach is the modification of records from the table by searching the first occurrence of the key term from the table and keep on selecting the records to include until the end of the table.

### 2. Indexed entity approach for modification (IEM approach)

A lookup table is created containing the identifiers of all the entities in the main table. Each record in the main table is indexed and similar names have the indices aggregated in the lookup table. Data is then modified with the aid of the lookup table. This is similar to IER approach.

# Deletion

## A. Maintenance of historical data

### 1. Transient deletion

Transient deletion is the removal of data, erasing it permanently. This is used when history is not important according to the database engineer.

### 2. Periodic deletion

Periodic deletion is the addition of data instead of the actual deletion of data and is labeled the updated value. The only value modified is the marker column indicating it is the current value.

Example:

Online assignment submission table

| Student name | Submitted | Timestamp | Action | Currency |
|---|---|---|---|---|
| Cruz, Michael T. | None | NULL | Not Submitted | Current |

And when the assignment was submitted, deleted, and then was resubmitted,

| Student name | Submitted | Timestamp | Action | Currency |
|---|---|---|---|---|
| Cruz, Michael T. | None | NULL | Not Submitted | NULL |
| Cruz, Michael T. | file(1).txt | 08-23-18 23:58:40 | Submitted | NULL |
| Cruz, Michael T. | file(1).txt | 08-23-18 23:59:54 | Deleted | NULL |
| Cruz, Michael T. | file(2).txt | 08-24-18 24:05:23 | Submitted | Current |

Here, it shows the multiple attempts of Michael to submit an assignment. **In a way, this entails that periodic deletion is essentially the same as periodic modification.**

## B. Deleting based on key term

### 1. Exhaustive search approach for deletion (ESD approach)

Exhaustive search approach is the deletion of records from the table by searching the first occurrence of the key term from the table and keep on deleting the records that match the key term until the end of the table.

### 2. Indexed entity approach for deletion (IED approach)

A lookup table is created containing the identifiers of all the entities in the main table. Each record in the main table is indexed and similar names have the indices aggregated in the lookup table. Data is then deleted with the aid of the lookup table. This is similar to IER and IEM approach.

# Database industrial techniques

## Updating/syncing of data warehouse

1. **Entirety syncing technique**

    Entirety syncing technique is the deletion of currently stored data and addition of all data from the first record to the last record (a combination of old data and new data). This is an acceptable technique when dealing with small datasets. Its disadvantage would be that it takes a significant amount of time to delete everything and add again what has already been added previously just to accommodate the new data. (Data Warehouse and Business Intelligence, 2014).

2. **Match syncing technique**

    Match syncing technique is the cross-matching between the stored data and the container that holds the combination of old and new data. If the old record exists in the stored data, move to the next record. If a record doesn't exist in the stored data, then add the record.

3. **Last sync pick-up technique (LSP technique)**

    Last sync pick-up technique involves checking for the timestamp of the last record of the stored data when execution is done daily, weekly, etc. Add the records that occur after the timestamp for that day or week. (Data Warehouse and Business Intelligence, 2014).

    Example:

    (1) *Dataset until yesterday $[d_1, d_2, d_3, ... d_n]$

    Data warehouse sync $[d_1, d_2, d_3, ... d_n]$ (base)

    (0) *Dataset until today $[d_1, d_2, d_3, ... d_n, d_{n+1}]$

    Data warehouse sync $[d_1, d_2, d_3, ... d_n]$ (base) + $[d_{n+1}]$ (data for today)

4. **Offset last sync pick-up technique (OLSP technique)**

    Last sync pick-up technique is only applicable in an *ideal* situation. If syncing is performed daily or weekly, there may come a time that the scheduled sync does not proceed due to power outage, internet inaccessibility, etc. Offset LSP technique provides

a workaround by introducing an *offset*. If the syncing is done daily, provide x number of days as an offset just in case it doesn't run for a day or more than that. If there are overlaps, delete the duplicates that would be generated.

Example:

> Set offset as 3 days
> (3) *Dataset until 3 days ago $[d_1, d_2, d_3, ... d_n]$
> Data warehouse $[d_1, d_2, d_3, ... d_n]$ (base)
>
> (2) *Dataset until 2 days ago $[d_1, d_2, d_3, ... d_n, d_{n+1}]$
> Data warehouse sync **doesn't** occur $[d_1, d_2, d_3, ... d_n]$ (same, no change)
>
> (1) *Dataset until yesterday $[d_1, d_2, d_3, ... d_n, d_{n+1}, d_{n+2}]$
> data warehouse sync **doesn't** occur $[d_1, d_2, d_3, ... d_n]$ (same, no change)
>
> (0) *Dataset until today $[d_1, d_2, d_3, ... d_n, d_{n+1}, d_{n+2}, d_{n+3}]$
> Data warehouse sync **occurs** $[d_1, d_2, d_3, ... d_n]$ (base)
> $+ [d_n]$ (duplicate of data already in the warehouse since offset is 3 days)
> $+ [d_{n+1}]$ (data for 2 days ago)
> $+ [d_{n+2}]$ (data for yesterday)
> $+ [d_{n+3}]$ (data for today).

The duplicate $d_n$ is then removed to preserve data.

# Data warehouse architecture

The process of extracting data from source systems and bringing it into the data warehouse is commonly called ETL, which stands for extraction, transformation, and loading. Note that ETL refers to a broad process, and not three well-defined steps. The acronym ETL is perhaps too simplistic, because it omits the transportation phase and implies that each of the other phases of the process is distinct. Nevertheless, the entire process is known as ETL. (Oracle, n.d.).

A data warehouse provides generalized data in multidimensional view. A data warehouse also provides Online Analytical Processing (OLAP) tools. These tools help in interactive and effective analysis of data in a multidimensional space. This analysis results in data generalization and data mining. (Tutorialspoint, n.d.).

There are five common architecture models for data warehouses namely: (1) two-level architecture, (2) independent data mart architecture, (3) dependent data mart and operational data store architecture, (4) Logical data mart and active warehouse architecture, and (5) three-layer architecture.

1. **Two-level architecture**

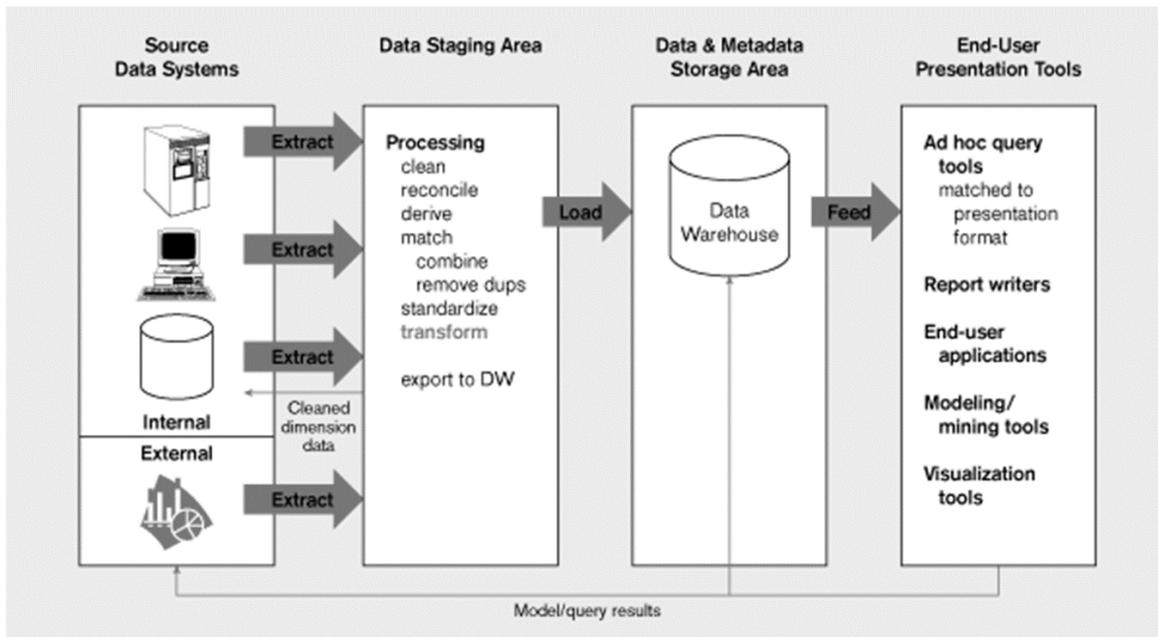

Figure 1.1 a two-level architecture

## 2. Independent data mart architecture (IDT architecture)

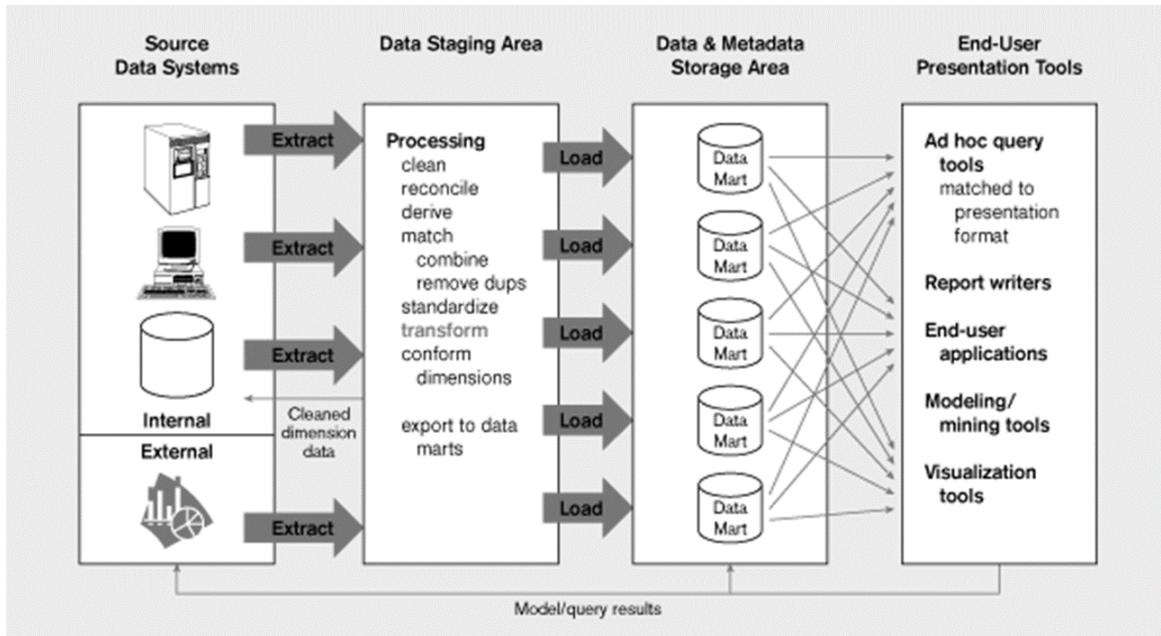

Figure 1.2 an IDT architecture

## 3. Dependent data mart and operational data store architecture (DDT architecture)

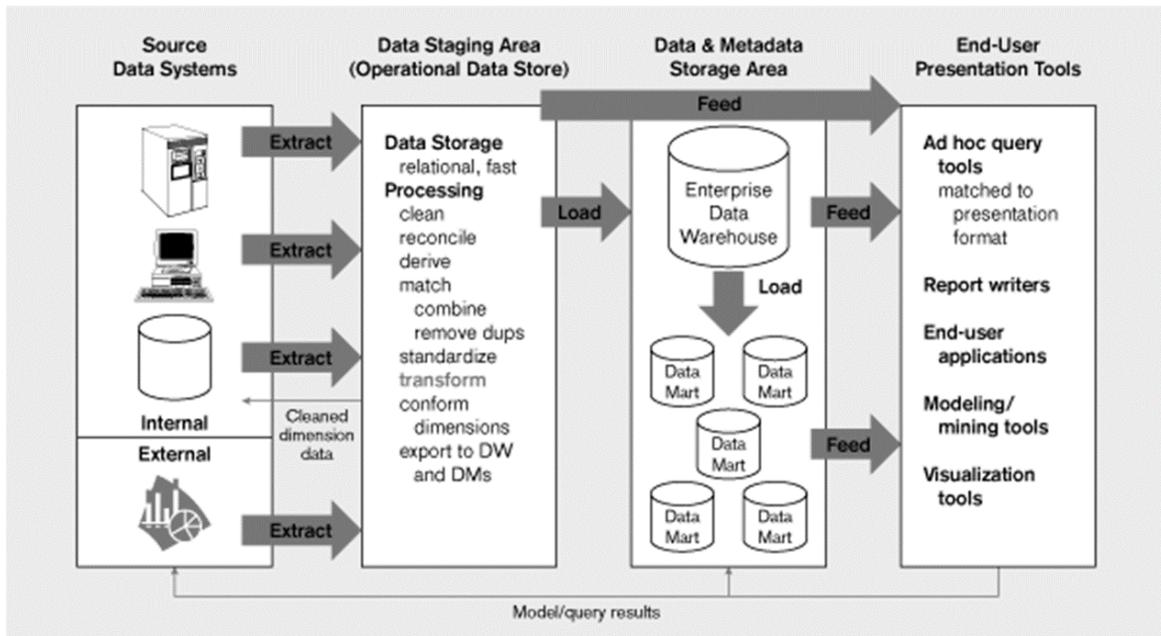

Figure 1.3 a DDT architecture

## 4. **Logical data mart and active warehouse architecture (LDM architecture)**

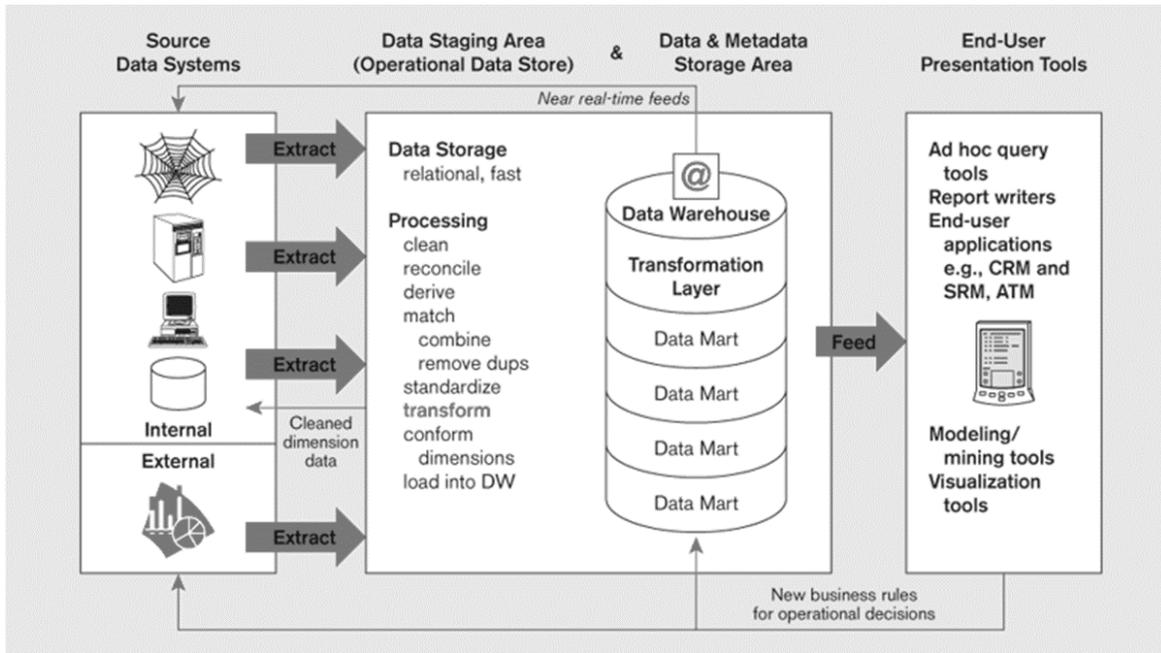

Figure 1.4 an LDM architecture

## 5. **Three-layer architecture**

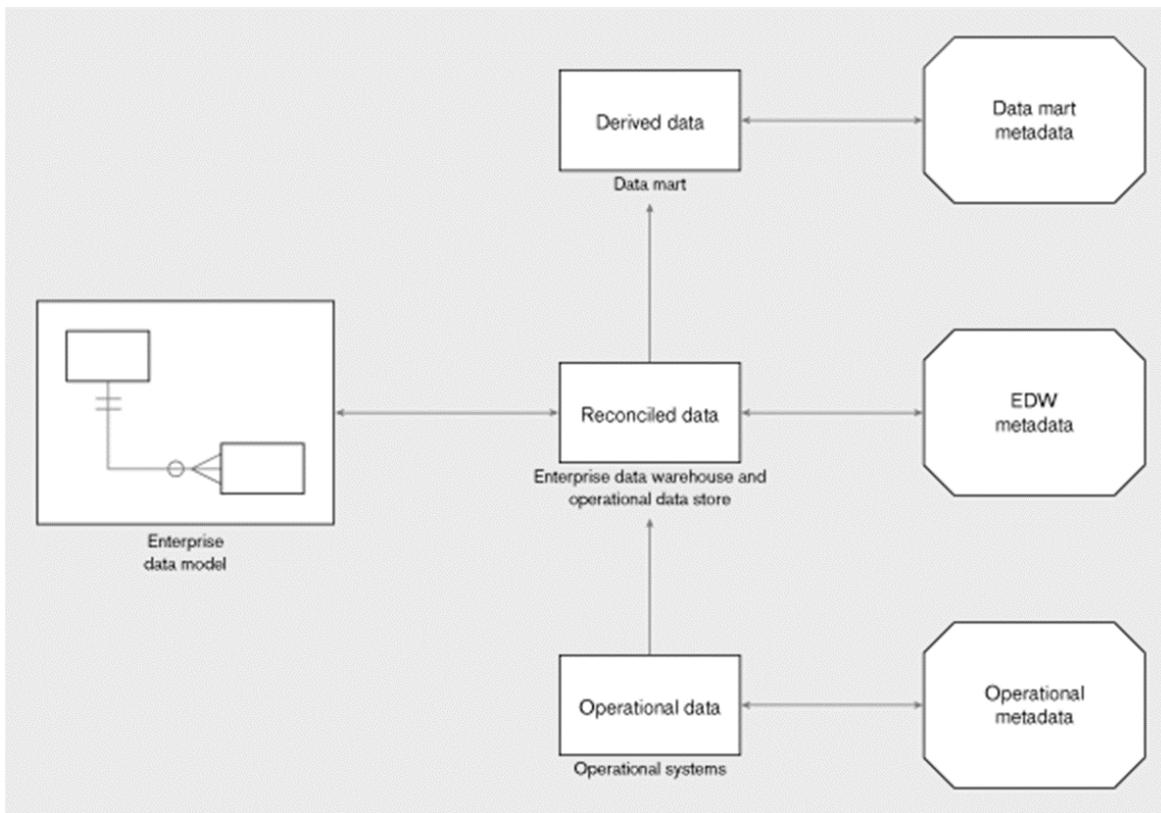

Figure 1.5 a three-level architecture

## Execution techniques

Now that the operation approaches and industrial techniques have been discussed, it is important to tackle the different execution techniques for the SQL query. Some of these techniques are as follows:

(1) In-editor manual execution
(2) Semi-automated scripted execution
(3) Semi-automated bashed execution
(4) Hybridized semi-automated execution
(5) Automated execution

### 1. In-editor manual execution

In-editor manual execution refers to the execution of an SQL query via by-hand input in an SQL editor. This is an acceptable technique if the query is ran extremely seldom or if the query is a one-time execution. Also, this is only ideal for short queries.

### 2. Semi-automated scripted execution

Semi-automated script execution refers to the execution of an SQL query via scripts like the built-in SQL script feature, or coding one in Python, Perl, Java, etc. The query itself is located inside the script file. Note that not all languages are packaged with SQL querying capabilities and libraries might be needed to achieve such task.

### 3. Semi-automated bashed execution

Semi-automated bashed execution refers to the execution of an SQL query via bash files (Mac OS/Linux) or batch files (Windows). The query itself is saved inside a .sql file and accessed by the bash or batch file through SQL server command line tools like sqlcmd. This makes the execution of the query easier with a double-click of the batch file per execution.

### 4. Hybridized semi-automated execution

Hybridized semi-automated execution refers to the execution of an SQL query via bash files or batch files that contain commands to run a script file. This combines the logic of the second and third technique. The SQL query is located inside the script file.

5. **Automated execution**

Automated execution refers to the execution of an SQL query via task scheduling manager. This takes a script file and executes it depending on the schedule that has been set. Some task scheduling managers for Windows are: Windows Task Scheduler (XP, 7), AT command, schtasks (8, 10), and Scheduled Tasks Cmdlets in Windows Powershell. For Mac OS and Linux, the most popular is the crontab—a configuration file that contains a cron table where cron jobs are placed.

**Cron**

People who set up and maintain software environments use cron to schedule jobs like commands or shell scripts to run periodically at fixed times, dates, or intervals. It typically automates system maintenance or administration—though its general-purpose nature makes it useful for things like downloading files from the Internet and downloading email at regular intervals. (Indiana University Knowledge Base, 2015).

Each line of a crontab is a task. It is important to understand its format to utilize it efficiently and properly. The format is as follows:

```
# ┌─────────────── minute (0 - 59)
# │ ┌───────────── hour (0 - 23)
# │ │ ┌─────────── day of the month (1 - 31)
# │ │ │ ┌───────── month (1 - 12)
# │ │ │ │ ┌─────── day of the week (0 - 6) (Sun-Sat
# │ │ │ │ │
# │ │ │ │ │
# │ │ │ │ │
# * * * * * ← This is the command to be executed.
```

| * any value | , value list separator | - range of values | / step values |
|---|---|---|---|

To try out simulations of cron expressions visit https://crontab.guru which is a free online tool for cron schedule expressions.

# About the author

I am a BS Computer Science student from the University of the Philippines Manila. I have also received training and industry experience on database management, data science and analytics, and machine learning. I wrote this peer-reviewed discussion paper so that I may share what I know from the past years as someone who strongly likes to work with data.

# Acknowledgment

I would like to thank my database professor in the University of the Philippines Manila, Sir Marbert Marasigan, for sparking the interest in me to delve into databases. You have imparted knowledge that I would carry over to my career and for the rest of my life. I am also super grateful to Sir Joseph Ramos and Sir JM Patiño for sharpening my database skills—especially on the industrial techniques I have discussed—during my internship at PLDT Global Corporation. I'd also like to thank my go-to professors anytime I want to share anything I've done in the field of computer science. Thanks for peer-reviewing this discussion paper, Sir Ruahden Dang-awan and Sir Berwin Yu. Last but definitely not the least, I am very appreciative for the motivation given to me by my friends in the University of the Philippines and the University of Sto. Tomas to engage the arts and sciences of databases.

# Dedication

This is for the devoted database architects and database engineers who want nothing but the best implementation of their operations on databases, for the students who want to pursue the intricacies of database management, and for the professors who aim to integrate the industrial methodologies regarding databases inside the classroom.